\documentclass[12pt,a4paper]{iopart}
\pdfoutput=1
\usepackage{graphicx}
\usepackage{upgreek}
\renewenvironment{pmatrix}[0]{\left(\begin{array}{*{20}{c}}}{\end{array}\right)}
\usepackage{mathrsfs}
\usepackage[margin=2.5cm]{geometry}
\newcommand{\msin}[0]{{\mathbf{s}}}
\newcommand{\mcos}[0]{{\mathbf{c}}}
\begin{document}
\title[Piezo films with adjustable anisotropic strain]{Piezo films with adjustable anisotropic strain for bending actuators with tunable bending profiles}
\author{Matthias C Wapler, Moritz St\"urmer, Jens Brunne and Ulrike Wallrabe}
\address{Laboratory for Microactuators, Department of Microsystems Engineering, \\
University of Freiburg, 79110 Freiburg, Germany}
\ead{wallrabe@imtek.uni-freiburg.de}
\begin{abstract}We present a method to produce in-plane polarized piezo films with a freely adjustable ratio of the strains in orthogonal in-plane directions. They can be used in piezo bending actuators with a tunable curvature profile. 
The strains are obtained as mean strains from a periodic polarization pattern produced by a suitable doubly interdigitated electrode structure. This mechanism is demonstrated for several examples using PZT sheets.
%
%
%
We further discuss how this tuning and the parameters of the electrode layout affect the overall magnitude of the displacement.\end{abstract}
\submitto{\SMS}
\section{Introduction} 
Piezoelectric materials have been widely used since the middle of the $20^{th}$ century in a myriad of actuators in many applications ranging from wavefront correction in astronomical telescopes (see e.g. \cite{astronomy} for a review) to fuel injectors in modern combustion engines (e.g. \cite{piezoinject} and citing papers).
A large part of these applications use piezo bending actuators where piezo films are joined with passive layers or differently polarized piezo films, recently in metal laminates (``THUNDER'', \cite{thunderpat,thunderpaper}), fiber laminates (``LIPCA'' \cite{lipca}) or functional gradient materials (``RAINBOW'' \cite{rainbow}). While in-plane polarized materials have been used just as long  (see e.g. \cite{adlerpat}), in the meanwhile most activity was based on out-of plane, i.e. transversely, polarized materials, where the electric field is applied using surface electrodes on either side of the film. In-plane polarized films with interdigitated electrodes (IDE), as they are used in surface acoustic wave filters (e.g. \cite{SAW}), have recently gained interest because of their larger displacements \cite{piezoefficiency} and their use in active fiber composites \cite{piezofibers} -- flexible laminates of 
piezo fibers. 
Generically, the bending profile is given by mechanical constraints, such as the boundary conditions, and the type of polarization chosen: Transversely polarized films will yield spherical displacements, in-plane polarizations will give saddle-shaped displacements, and active fiber composites give displacements along the direction of the fibers. Attempts to obtain other displacements are usually limited in actuation range and flexibility and involve a complicated set of many electrodes and control voltages \cite{flexbendmirr}.

In this paper, we demonstrate how to use the full potential of the anisotropy of the transverse and longitudinal piezoelectric effect in order to create geometrically isotropic and homogeneous piezo composites with freely adjustable bending profiles. 

To do so, we develop a doubly interdigitated electrode (DIDE) structure that gives a fine piecewise in-plane polarization pattern in different directions with the desired macroscopic ratio of the mean strains in the plane. In section \ref{theory}, we will present a simple model that gives the appropriate displacement parameters independent of the specific structure of the layering and the material properties. We will use this to study how the geometric parameters of the electrode structure affect the strain ratio and total strain in the plane. The strain in the plane predicted by the theory is compared in the experimental section to the displacement profile of a bending actuator consisting of a piezo sheet and a passive glass layer. The rapid prototyping process and material properties are described in section \ref{process}, and the displacements were measured using a laser profilometer setup and analyzed as we describe in section \ref{characterization}. In \ref{deriv}, we derive the mean strain of the piezo sheets from first principles, which lays the theoretical foundation for this paper.
%
\section{Theory}\label{theory}
Let us assume that we have a thin piezo layer with anisotropic strain, parametrized by $s_{xx}$ and $s_{yy}$, that is bonded on top of a thin substrate. When the piezo expands or contracts in either direction, the system will bend with an out of plane displacement $z$. 
As we are not interested in the details of the mechanics, we can parametrize the effects of forces and of the details of the stacking by some parameter $n$:
\begin{equation}\label{bendeq}
z(x,y) \ = \ z_0 - \frac{s_{xx}}{2 n}(x-x_0)^2- \frac{s_{yy}}{2 n}(y-y_0)^2 \ .
\end{equation}
This is just the displacement of a generic linear system with parabolic profile. It actually corresponds purely geometrically to the limit of small displacements of a system that has a passive plane and a strained plane with strains $s_{xx}, \, s_{yy}$, separated by a distance $n$.

For an ordinary transversely polarized ``$d_{31}$'' actuator, $s_{xx}=s_{yy}=d_{31} E_z\sim -\frac{d_{33}}{2}E_z$, so we will have a spherical profile. 
For an in-plane polarized ``$d_{33}$'' actuator for which we take $x$ as the polarization direction, we have $s_{xx}=d_{33}E_x$ and $s_{yy}=d_{31}E_x\sim - \frac{1}{2}s_{xx}$. Hence, it will form a saddle-shaped displacement.

\subsection{Basic Principle}\label{basic}
In this paper, we want to go beyond these restrictions and create a free bending profile. 
Rather than polarizing the system only in a single direction, we polarize it piecewise in different directions. We will actuate it with the same electrodes that we use for the initial polarization, such that locally, we will have an expansion $d_{33} E$ parallel to the field, and a contraction $d_{31} E$ transverse to the field -- provided the system behaves linearly. The field is always taken positive in the direction of the polarization. If the actuation field is not the same as the initial polarizing field, one has to consider the projection onto the remanent polarization. 

In  \ref{deriv}, we derive the mean strain from first principles, and if we assume that $d_{31} = -\frac{1}{2}d_{33}$, the resulting mean strain over a patch is 
\begin{eqnarray}\label{meanstrains}\nonumber
s_{xx}\sim d_{33} \left( \left<\frac{E_x^2}{|E|}\right>-\frac{1}{2}\left(\left<\frac{E_y^2}{|E|}\right> +\left<\frac{E_z^2}{|E|}\right>\right)\right) \ \ \mathrm{and}\\ s_{yy}\sim d_{33} \left(\left<\frac{E_y^2}{|E|}\right>-\frac{1}{2}\left(\left<\frac{E_x^2}{|E|}\right> +\left<\frac{E_z^2}{|E|}\right>\right) \right) \ .
\end{eqnarray}
This is also the mean strain over the whole sheet, provided all patches are equal.
The average is taken over the area, and we included the out-of-plane component of the field that will never vanish completely if we use surface electrodes. Strictly speaking, we consider here the field relative (and parallel) to the direction of the remanent polarization.
This allows us to tune the bending radii in both directions. For example $\left<\frac{E_x^2}{|E|}\right>=\frac{1}{2}\left(\left<\frac{E_y^2}{|E|}\right> +\left<\frac{E_z^2}{|E|}\right>\right)$ and $\left<\frac{E_y^2}{|E|}\right>=\frac{1}{2}\left(\left<\frac{E_x^2}{|E|}\right> +\left<\frac{E_z^2}{|E|}\right>\right)$ make $s_{xx}$ or $s_{yy}$, respectively, vanish, resulting in a bending profile in only one direction and a second flat direction and $\left<\frac{E_x^2}{|E|}\right>=\left<\frac{E_y^2}{|E|}\right>  $ set $s_{xx} = s_{yy}$, resulting in a spherical displacement. 

\begin{figure}
\begin{center}\raisebox{0.25\textwidth}{\bf a)}\includegraphics[width=0.4\textwidth]{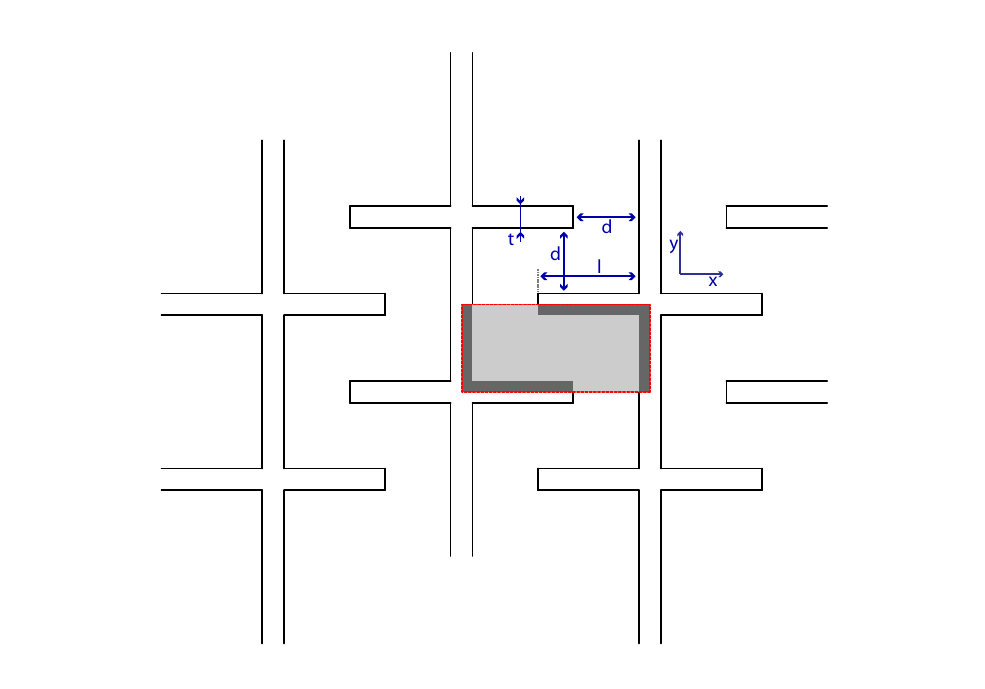} \raisebox{0.25\textwidth}{\bf b)}\includegraphics[width=0.3\textwidth]{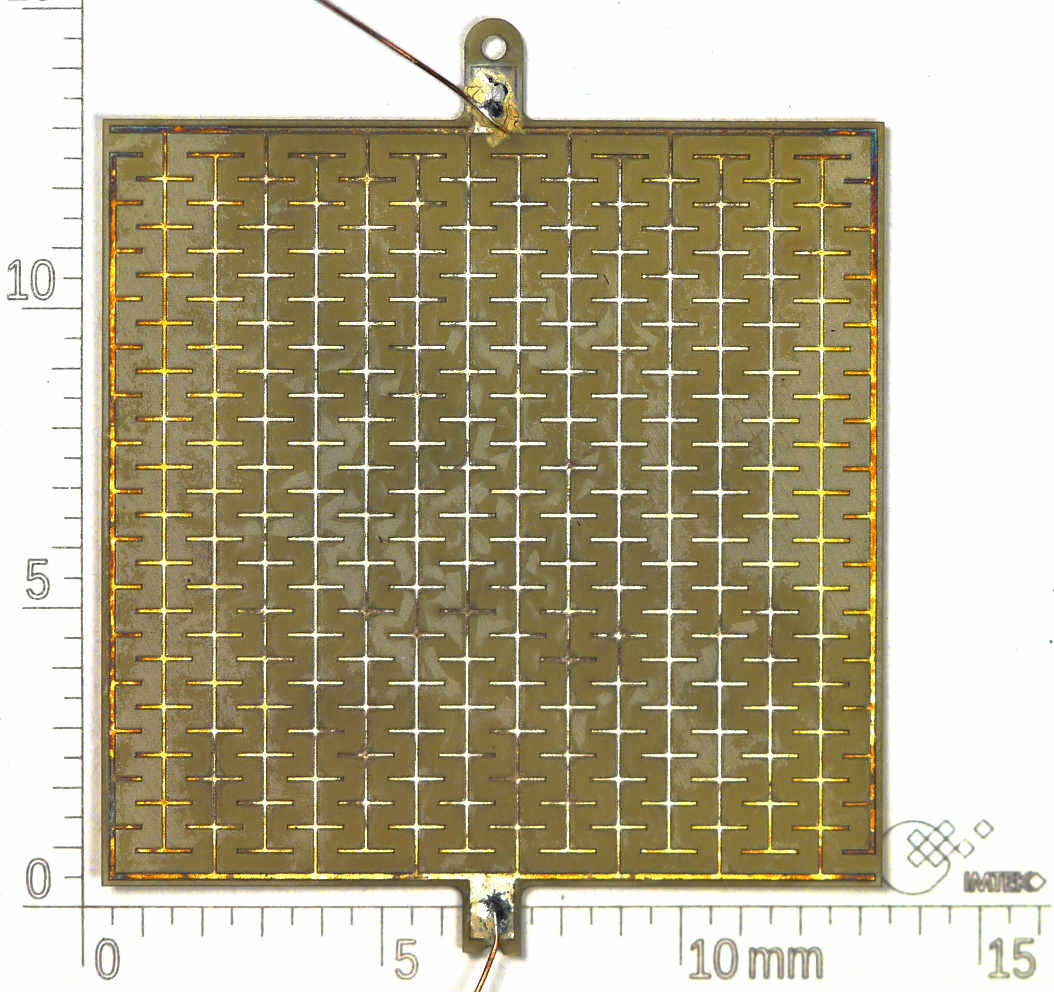}\\
\raisebox{0.266\textwidth}{\bf c)}\includegraphics[width=0.61\textwidth]{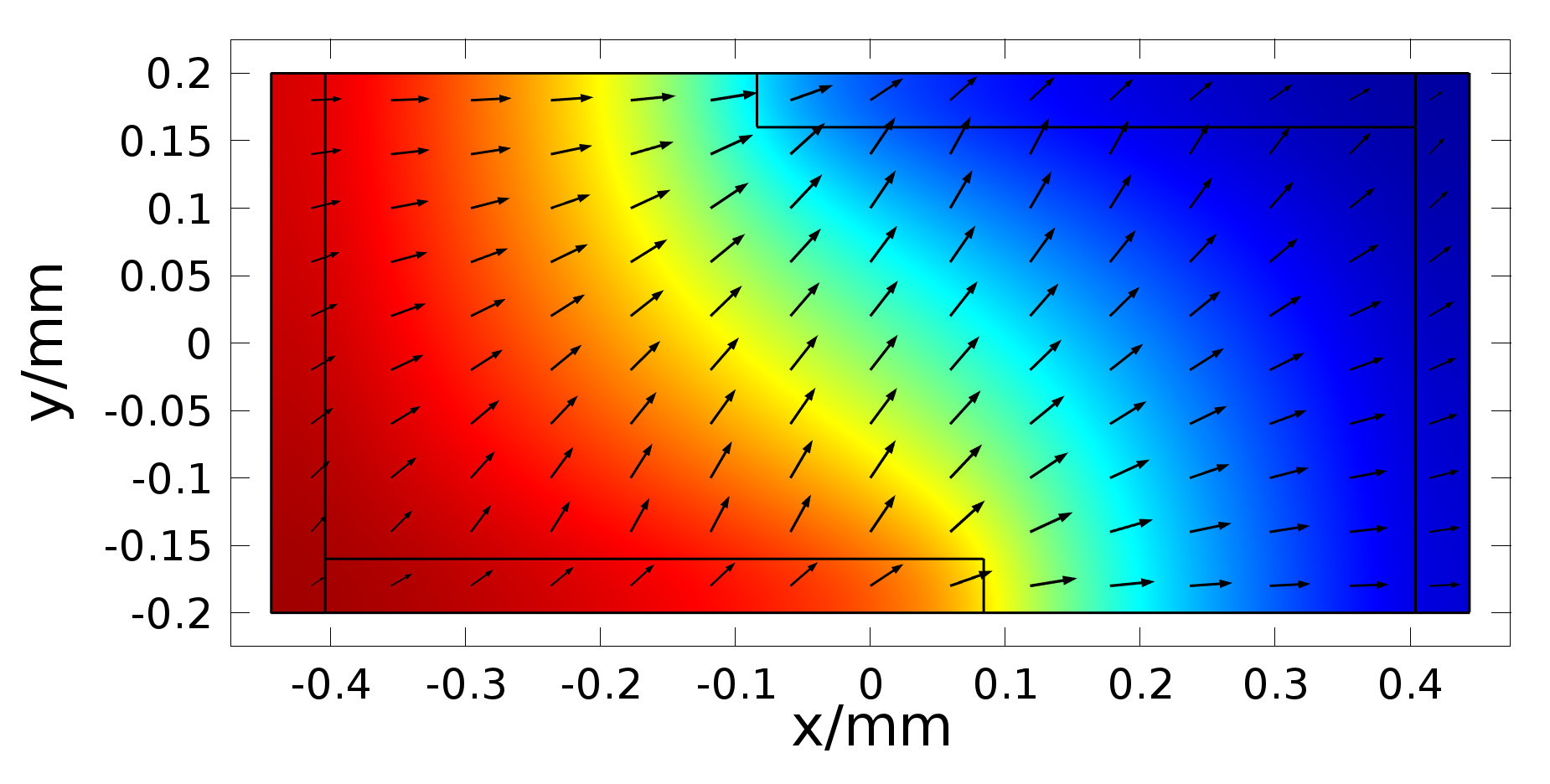}
\includegraphics[width=0.34\textwidth]{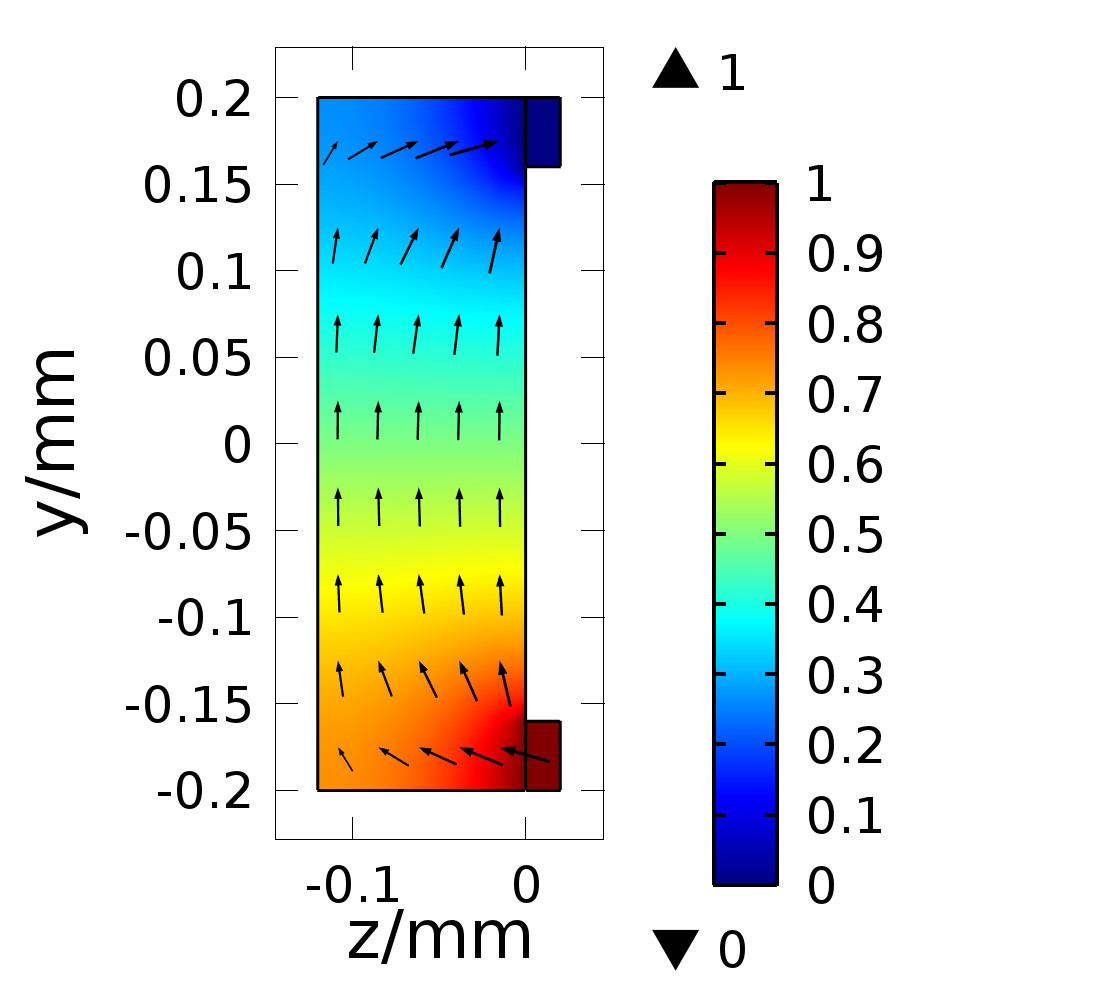}\end{center}
\caption{a) Electrode layout and b) image of the piezo sheet. c) Simulation of the electric potential in the unit cell (highlighted above) in a single-sided design, normalized to $1$. Top view of a plane in the middle of the sheet, at $z=-60\, \upmu \mathrm{m}$ and cut at $x=0$, viewed from the left. The black lines indicate the electrodes on top.}\label{layout}
\end{figure}
To realize such flexible bending profiles, we consider a thin piezo sheet with thickness $h$ on which we have a doubly interdigitated electrode design shown in fig. \ref{layout}a that consists of electrodes with width $t$ and separation $d$ and a length $l$ of the secondary fingers. This finger length can then be used to tune the bending profile. We will consider designs with electodes only on one side of the piezo and also double-sided designs in which electrode patterns are aligned exactly on top of each other on both sides. This ensures that there is no additional transverse polarization and the field geometry is given by a mirroring of a single sided electrode pattern with half the piezo thickness.

To illustrate the tuning of the strains, we consider a very naive geometric model in which the electrode width vanishes, $t\rightarrow 0$. Let us assume that in the triangles between the tips of the secondary fingers and the main fingers, the field goes entirely in the $x$ direction and in the parallelogram between the secondary fingers, the field goes entirely in the $y$ direction, both with strength $E = \frac{U}{d}$.
Then, geometry tells us that $\left< \frac{E_y^2}{|E|} \right> \sim \frac{l}{d}\left< \frac{E_x^2}{|E|}  \right> $.

To obtain more accurate values, we perform a three-dimensional  finite element simulation of the unit cell indicated in fig. \ref{layout}a.
We see in the right plot of fig. \ref{layout}c that there is a significant transverse electric field $E_z$ near the electrodes and that there are inefficiencies with large fields below the electrodes and in the left plot at the outer (convex) corners and vanishing fields at the inner (concave) corners of the electrodes. This means that we may get saturation in some parts of the material while other regions of the material are not fully used.
%

As our design has been chosen to facilitate straightforward rapid prototyping and characterization, it is far away from the optimal design. We conjecture the latter to be the the two-dimensional reduction of a (D)IDE design in the limit where simultaneously $h\rightarrow 0$ and $t\rightarrow 0$ or first $h\rightarrow 0$ and then $t\rightarrow 0$). This model is scale-invariant and only depends on the ratio $l/d$. How one may approach this optimum has been discussed for conventional IDE configurations \cite{piezoefficiency}, and we will illustrate this for the present DIDE design later in this paper. Still, such an orthogonal IDE design is not optimal as one looses always some factor in eq. \ref{meanstrains} due to Pythagoras' theorem. As the optimum, we hence conjecture a hypothetical design with a homogeneous strength $|E|$ of the electric field that varies abruptly by $90^\circ$ in small patches that are polarized either in the $x$ or in the $y$-direction. Then, $\left<E_x\right>+\left<E_y\right> = |E|$ and $\left<E_z\right>=0$, such that this configuration can be described by a parameter $q \in [ 0,1 ] $:
\begin{equation}
\left<E_x\right> \ = \ |E| \, q \ \ \mathrm{and} \left<E_y\right> \ = \ |E| \, (1-q) \ .
\end{equation}
Our naive geometrical model above would be such a configuration with $q = \frac{l}{l+d}$.
Using a different parameter, $\theta = \tan^{-1}\left( \frac{1-3q}{3q-2}\right)$ with the branch cut such that $\theta \in \left[\tan^{-1}\frac{1}{2}, \tan^{-1}2 \right] \simeq [-0.464, 2.034] \simeq \left[\frac{\pi}{4} -1.249,\frac{\pi}{4} +1.249\right]$,
 the strain is described by
\begin{equation} \label{hypothet}
s_{xx} \ = \ d_{33}|E| \frac{\cos\theta}{2\left(\sin\theta + \cos\theta \right)} \ \ \mathrm{and} \ s_{yy} \ = \ d_{33}|E| \frac{\sin\theta}{2\left(\sin\theta + \cos\theta \right)} \ .
\end{equation}
There, we can easily identify the ratio  $\frac{s_{yy}}{s_{xx}} \, = \, \tan \theta$ and the total strain. 

\subsection{Simulations}\label{simsec}
In this section, we will study how the shape of the displacement depends on the parameters of the geometry of the DIDE layouts and how efficient they are, i.e. how large is their displacement for fixed electric field and material properties. Assuming full mechanical and electrical linearity, the simulation is just a straightforward electrostatic simulation of the volume of one patch of the DIDE pattern. The symmetry and periodicity of the DIDE pattern and the further assumption that the dielectric constant of the piezo material is very large, $\varepsilon_r \gg 1$\footnote{This is satisfied by PZT materials with $\varepsilon_r$ of the order of 1000.}, imply the boundary condition $\vec{E} \cdot \hat{n} = 0$ for all surfaces of the volume, except for the electrodes where we apply the appropriate potentials. In other words: No electric field lines (or electric flux) leave the unit cell, except for the surface of the electrodes. The resulting field configuration is used to obtain the mean strains from eq. (\ref{meanstrains}).

To analyze this model in a scale-invariant fashion that does not depend on the material parameters, we describe the shape of the displacement with the parameters $s_{xx}/s_{yy}$ and $s_{yy}/s_{xx}$. A value of $1$ corresponds to a spherical displacement of the bending actuator, $0$ corresponds to one flat direction and $-1$ to a saddle function with equal (but opposite) curvature in both directions. To consider the magnitude of the displacement, we will use the ratio of $s_{xx}$ or $s_{yy}$, whichever is larger, to the hypothetically achieveable maximum strain $s_{max.} = d_{33} |E|$ at a nominal electric field strength $E=\frac{U}{d}$. The geometry will similarly be  parametrized by a ratio to parameters that we keep fixed, for example by the ratio $l/(d+t)$ in the case of varying $l$ and fixed $(d+t)$. 

\begin{figure}\begin{center}
{\raisebox{0.26\textwidth}{\bf a)}}\includegraphics[width=0.455\textwidth]{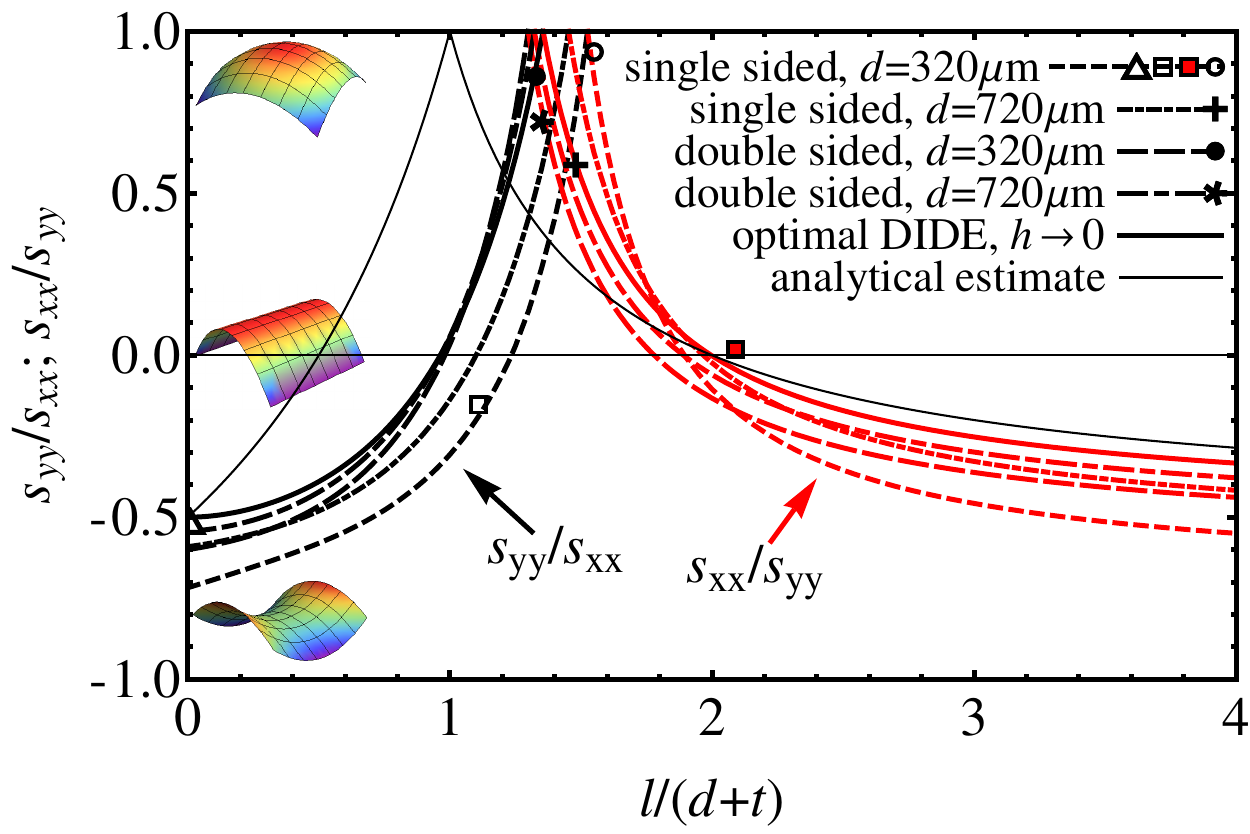}
{\raisebox{0.26\textwidth}{\bf b)}}\includegraphics[width=0.445\textwidth]{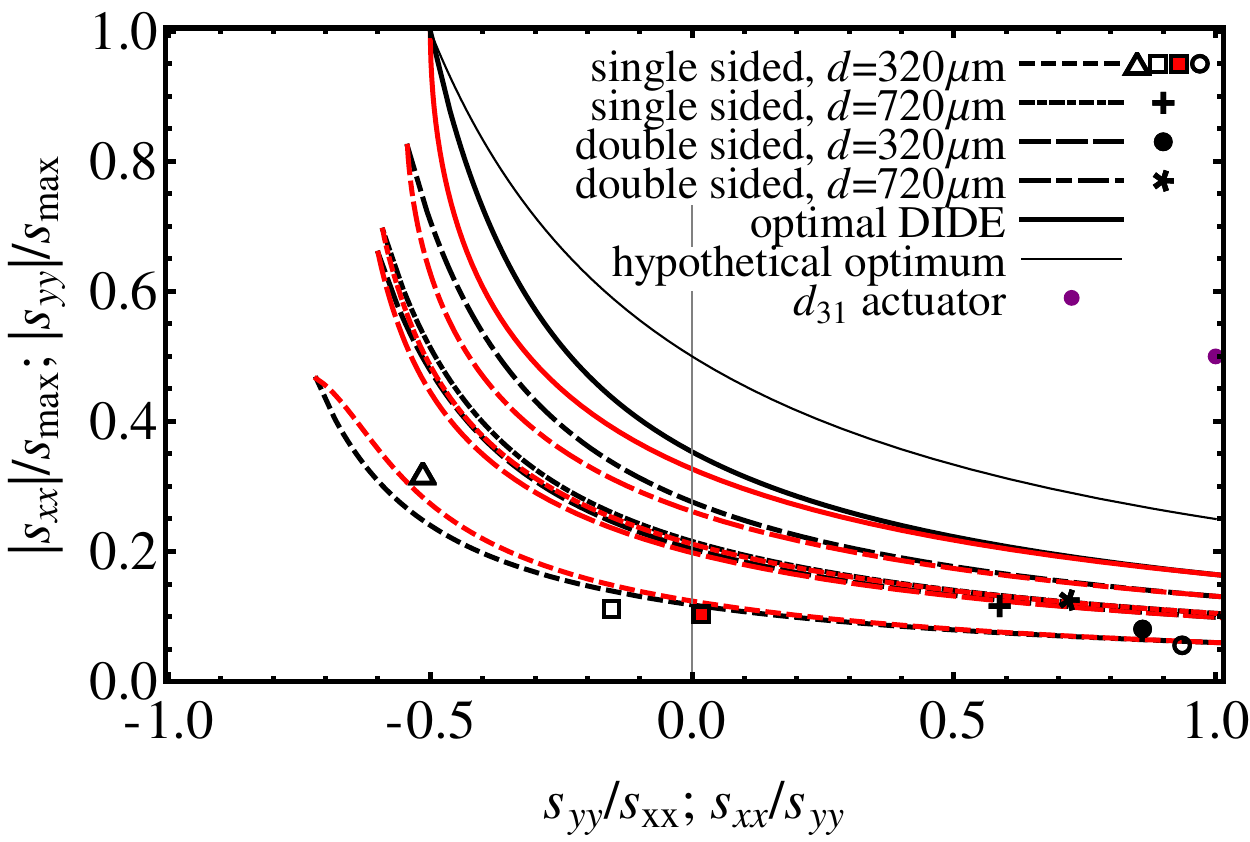}\\[5mm]
{\raisebox{0.26\textwidth}{\bf c)}}\includegraphics[width=0.45\textwidth]{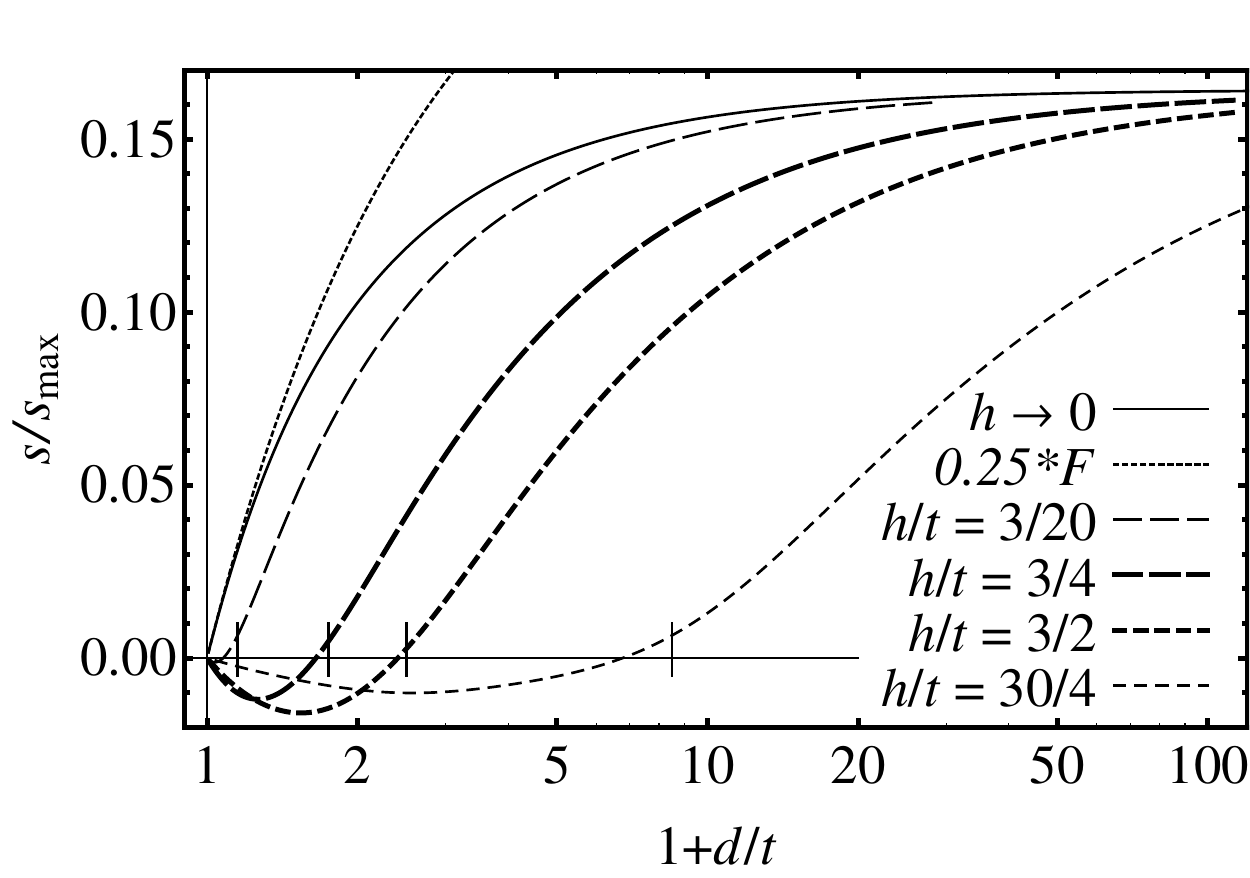}
{\raisebox{0.26\textwidth}{\bf d)}}\includegraphics[width=0.45\textwidth]{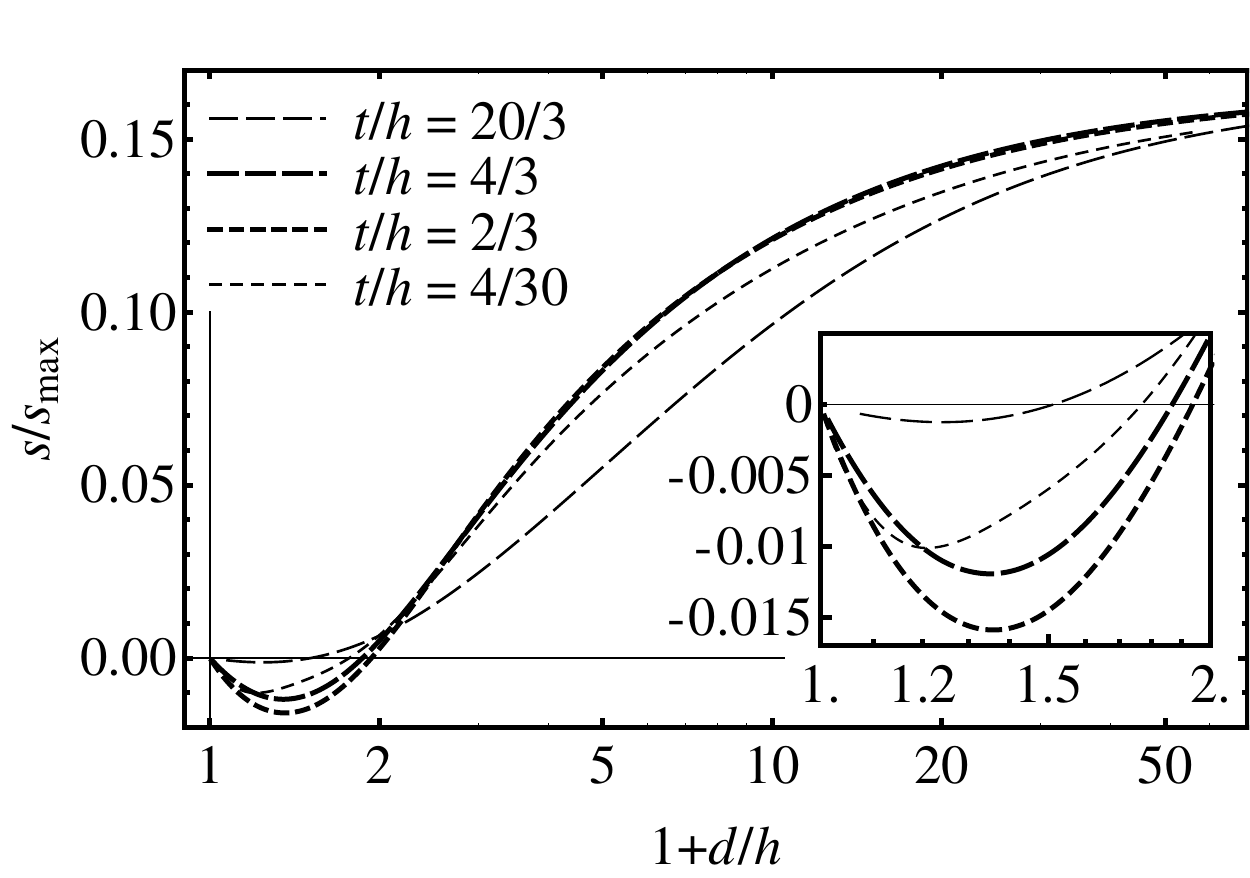}\end{center}
\caption{a) Shape of the displacement depending on the ratio of the finger length $l$ to the height of the unit cell $d+t$ for fixed $h=120 \, \upmu \mathrm{m}$ and $t=120 \, \upmu \mathrm{m}$ and various values of the electrode spacing $d$. The solid line shows an optimal DIDE design, the narrow line shows the analytical approximation of section \ref{basic}, and dashed curves describe the simulation results for the layouts of our prototypes. The black and red lines show $s_{yy}/s_{xx}$ or $s_{xx}/s_{yy}$, whichever is smaller in magnitude. \\
b) Displacement of the same actuators compared to the maximum displacement $s_{max} = d_{33} |E|$ as a function of the shape. The narrow line is the hypothetical optimum (\ref{hypothet}) and the symbols represent the measured data described in section \ref{characterization}. The triangle corresponds to the conventional IDE ($l=0$), the boxes to the unidirectional actuators and the circles, star and plus to the spherical (isotropic) displacement. \\
c) Strain of a spherical actuator for fixed electrode width $t$ for various film thicknesses $h$. The thick dotted and dashed lines correspond to the single- and double sided (which effectively has $h=60\, \upmu \mathrm{m}$) layouts of a) and b) and the thin dotted and dashed lines the narrow electrodes and thin sheets, respectively. The narrow vertical lines indicate $1, \ 1+4/30, \ 1+2/3, \ldots$, corresponding to the values of $h/t$. \\
d) Strain of a spherical actuator for fixed film thickness for various electrode widths. The inset shows the region of small $d/h$, where the actuator effectively contracts. }\label{lengthplot}
\end{figure}

In fig. \ref{lengthplot}a, we show the shape as a function of this dimensionless parametrization of the finger length $l$ for various values of the electrode separation $d = 320 \, \upmu \mathrm{m},\ 720\, \upmu \mathrm{m}$ and double- and single-sided electrodes. The piezo thickness is $h= 120 \, \upmu \mathrm{m} $ and the electrode width is $t =80 \, \upmu \mathrm{m} $. This will also be the dimensions in the experimental section. 


We find that $s_{yy}/s_{xx}$ starts at $l=0$ with $-1/2$ for the optimal DIDE design ($t\rightarrow 0$, $h\rightarrow 0$) as we expect from eq. (\ref{meanstrains}) since we assumed $d_{31} = -\frac{1}{2}d_{33}$. When the secondary fingers are slightly larger than the electrode gap,  $\frac{l}{d+t} \sim 1.4$, we have a sphere, and for long fingers we approach a saddle in the opposite direction (red lines). For the realistic layouts, we see that the double-sided electrode layouts are closer to the optimal DIDE and the single-sided layouts are further away but still follow the same behavior, so the ratio $\frac{l}{d+t}$ is a good universal parameter to describe the layouts. The deviations are most significant near $\frac{l}{d+t} \sim 1$, where the dependence on the finger length is also most critical. Interestingly, at small $\frac{l}{d+t}\ll 1$, the realistic DIDE layouts go beyond the limit of $s_{xx}/s_{yy} = - 1/2$; probably due to the transverse effect of the out-of-plane field.

As a measure for the displacement, we show in fig. \ref{lengthplot}b the larger of the linear strains $s_{xx}$ and $s_{yy}$, corresponding to the curvature of the minor axis of an elliptical or saddle-shaped bending actuator. 
We see that for all saddle configurations ($\frac{s_{xx,yy}}{s_{yy,xx}}<0$, the double indeces represent the two possible ratios, whichever is smaller in magnitude), the hypothetical optimal actuator described by (\ref{hypothet}) has a larger displacement than a conventional transversely polarized (``$d_{31}$'') piezo film, using $d_{31}= - \frac{1}{2} d_{33}$. Once we go into the region of ellipsoidal and spherical displacements ($\frac{s_{xx,yy}}{s_{yy,xx}}>0$), the optimal in-plane polarized piezo sheet has less displacement than the transverse actuator. Hence, the variable shape comes at the expense of a decreased displacement. The specific designs of our DIDE prototypes that we described above are significantly less efficient than the optimal design, with the single-sided electrode layout performing particularly poor. This is not unexpected as the transverse effect of $\left<E_z\right>$ counteracts the longitudinal effect that drives the displacement. 
This effect may also explain why for most realizable DIDE structures the red branch (large $\frac{l}{d+t}$, $|s_{yy}|>|s_{xx}|$) is more efficient than the black branch (small $\frac{l}{d+t}$, $|s_{xx}|>|s_{yy}|$), while for the optimal DIDE limit it happens the other way round. Small $\frac{l}{d+t}$ also have the advantage that the displacement is more 
homogeneous and there is less potential for rippling in the surface. As we can see from the displacement of the optimal DIDE layout, there is still room for improvements compared to our actuators. Coming very close to the hypothetical optimum (\ref{hypothet}) will be difficult, or even impossible as it is well above the optimal DIDE structure and it is difficult to imagine electrode configurations that have more abrupt changes of the field direction without locally reducing the electrode spacing (and thus reducing the possible operation voltages).
%
%
%
%
%

In fig. \ref{lengthplot}c, we study this further and look at the efficiency $\frac{s}{s_{max}} = \frac{s}{d_{33} |E|}$ at $s_{xx}=s_{yy}$ depending on the electrode spacing $d$ for different piezo thicknesses, scaled by the electrode width $t$ in a log-linear plot ($1+d/t$ to also look at $d\rightarrow 0$). As we saw before, a wider electrode spacing is more efficient, as are thinner piezo films (or double-sided electrodes, corresponding by symmetry to half the piezo thickness). We see that for electrode spacings less than the piezo thickness, the expansion from the in-plane field is counter-acted by the contraction due to the transverse field component. Only in the limit of extremely thin piezo films, the piezo area $A_{piezo}$ obstructed by the electrodes $A_{\mathrm{electrode}}$ becomes relevant, as we see from the dotted line that shows 0.25 (the efficiency of the hypothetical optimum) times the fraction $F = \frac{A_{\mathrm{piezo}}-A_{\mathrm{electrode}}}{A_{piezo}} = \frac{d}{d+t}$ of the free surface.

In particular if we look at fig. \ref{lengthplot}d where we take the piezo thickness $h$ as a scaling parameter,
we see that the relevant quantity that controls the effects over a large range of configurations is just the ratio of the electrode spacing to the piezo thickness, $d/h$. 
We see from the curves with different values of $d/t$ in the plot that the optimal electrode width is not infinitely narrow, but of the order of the piezo thickness.
This is because for very narrow electrodes, the efficiency decreases again as the field diverges near the narrow electrodes where there is a logarithmic drop in the potential. Hence, there is less field strength in the bulk of the material. Certainly this effect will change if we take saturation into account.

From these plots, we see that our conjecture for the optimal DIDE of taking $h\rightarrow 0$ and $t\rightarrow 0$ was indeed correct, provided $t$ is not first taken to zero. Furthermore, we see from the plots that realistic configurations with $d/h \sim 10 \ldots 20$ may achieve around 80\% to 90\% of an optimal DIDE actuator or 50\% to 60\% of the hypothetical optimum in the spherical case and more for the saddle-shaped displacements where the hypothetical optimum and the optimal DIDE meet.

We expect that the saturation effects will be apparent in the experimental section, where we test different electrode spacings and single- and double-sided electrode layouts corresponding to different effective film thicknesses. The effects should be most prominent for narrow electrodes, i.e. large spacings or thicknesses.
\section{Design and processing}\label{process}
To verify our model and the predictions from section \ref{simsec}, we produced prototypes with approximate square shapes around 13 to 14 mm.
We use commercial $ 120 \pm 20 \, \upmu \mathrm{m}$ thick PZT ceramic disks with a diameter of 25mm and $ 5\, \upmu \mathrm{m}$ thick silver electrodes on both sides, as they are used in acoustic transducers. The advantage of this material is a low $\mathrm{T_c}$ of $\sim 250 ^\circ \mathrm{C}$ that is due to doping with $\sim 2.7\% $ strontium, and a large coercive field strength of $\sim 950 \, \mathrm{V/mm}$ -- allowing for reliable depolarization at relatively low temperatures, good polarizability and a large range of actuation. We further roughly measured a coefficient $d_{31} = -2.7 \pm 0.3 \times 10^{-4} \mathrm{mm/kV}$ and a Young's modulus of $E \sim 50 \pm 8 \,  \mathrm{GPa}$ using simple bending actuator and cantilever setups.

As a passive layer, we use common microscope cover slides with $\sim 141\pm 2 \, \upmu \mathrm{m}$ thickness and $E = 67\pm 3 \mathrm{GPa}$. The bonding is performed with Araldite 2020, a transparent low-viscosity epoxy glue that is recommended by its manufacturer for glass and ceramic bonding. In principle, we could have created a passive layer inside the piezo by leaving an unstructured electrode on one side and thus introducing an equipotential plane with $s_{xx}=s_{yy} \sim 0$. As we would need to consider the $z$-dependence of the strain, this would, however, not allow for such a simple model as in sec. \ref{theory}. It would furthermore drive the field between the electrodes into the non-linear regime close to saturation at usual operating voltages and electrode geometries. 

For the fabrication of the prototypes, we use a simple rapid prototyping process: First, we structure the electrodes and cut the outer contour of the piezo sheets by ablation with a UV laser, leaving $700\, \upmu \mathrm{m} \times 1000 \, \upmu \mathrm{m}$ contact pads at the sides. As the laser intensity tends to fluctuate, this process also removes a few $\, \upmu \mathrm{m}$ of the PZT. Next, we remove the residues of the PZT and the silver in weak ultrasound in purified water with a dip of a few seconds in 25\% $\mathrm{HNO}_3$ at room temperature and subsequently dry the PZT sheet. This leaves an electrode width of $80\pm 20 \, \upmu \mathrm{m}$. The PZT is then depolarized at $420 ^\circ \mathrm{C}$, well-above the Curie temperature,  temporarily coated with  photoresist to prevent electric breakdown and subsequently polarized in-plane by applying a nominal field $U/d$ of $\sim 2\ \mathrm{kV/mm}$ via the electrode pattern for $\sim 10\, \mathrm{min.}$ to allow for sufficient creep close to saturation. After stripping the photoresist, the piezo sheets are glued to the 
passive layer. A 
finished piezo without passive layer is shown in fig. \ref{layout}b.
%

For the characterization, the samples are supported at their center by adhesion on a $4\, \mathrm{mm}\times 4\,\mathrm{mm}$ wide, $\sim 1.6 \, \mathrm{mm}$ thick piece of soft polyurethane with shore hardness A10. 
To measure the surface profile, the actuator is operated with a quasistatic triangular signal of $5 \, \mathrm{Hz}$ with offset and amplitude to give a maximum of $625 \mathrm{\ V/mm}$ (in the direction of the polarization) and a minimum of $-125 \mathrm{\ V/mm}$ (against the remanent polarization) which is much smaller than the coercive field strength. The surface is then scanned with a laser triangulation sensor with an averaging of three cycles at each point. 
All displacements are taken relative to the initial displacement, and values with more than $2\sigma$ deviation from the local mean displacement were discarded in order to eliminate excessive noise coming for example from contamination of the surface. Furthermore, the data from the outer  $\sim 250 \, \upmu \mathrm{m}$ wide region of the piezo was not considered in the data analysis to remove noise coming from the edge of the sheet.
\section{Results}\label{characterization}
\begin{figure}
\begin{center}{\includegraphics[width=0.24\textwidth]{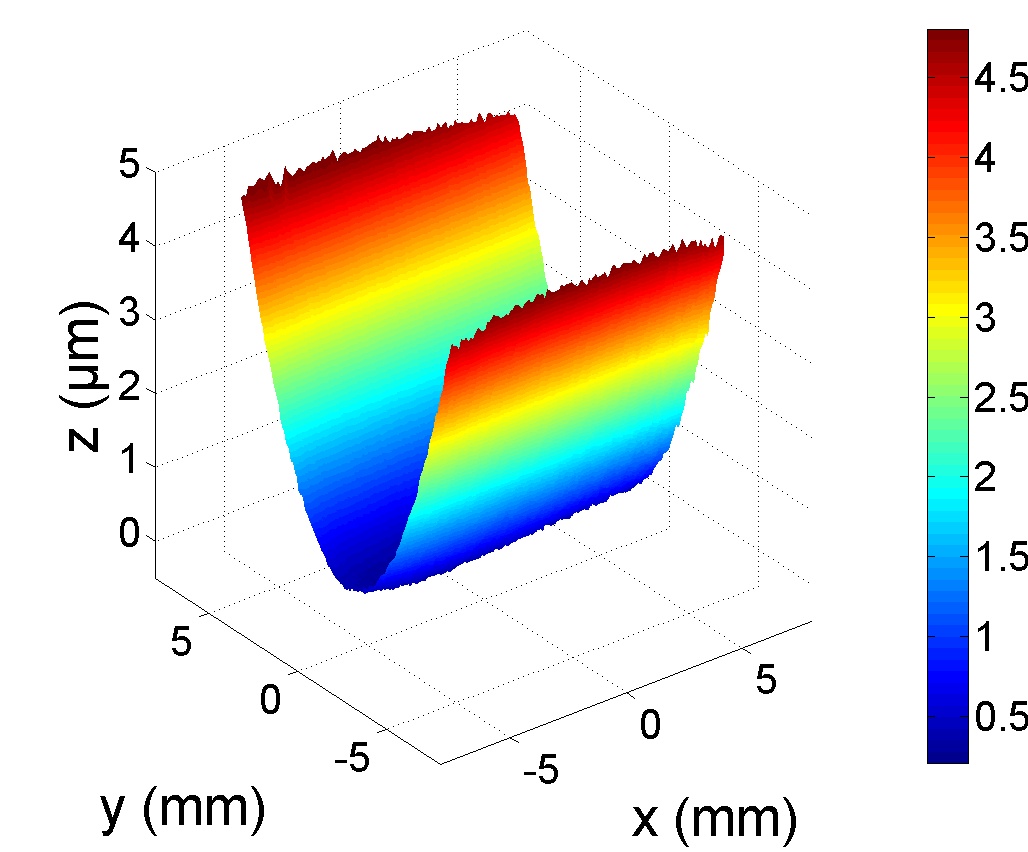}\hspace{0.001\textwidth}
\includegraphics[width=0.24\textwidth]{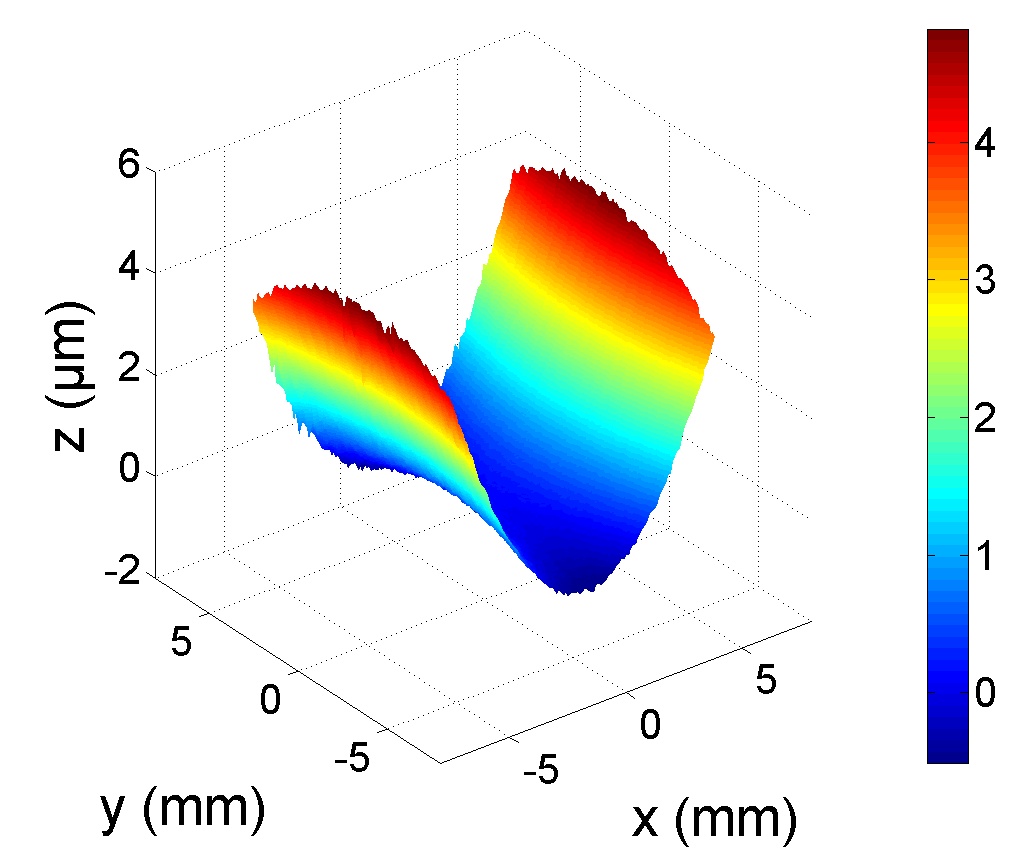}\hspace{0.001\textwidth}
\includegraphics[width=0.24\textwidth]{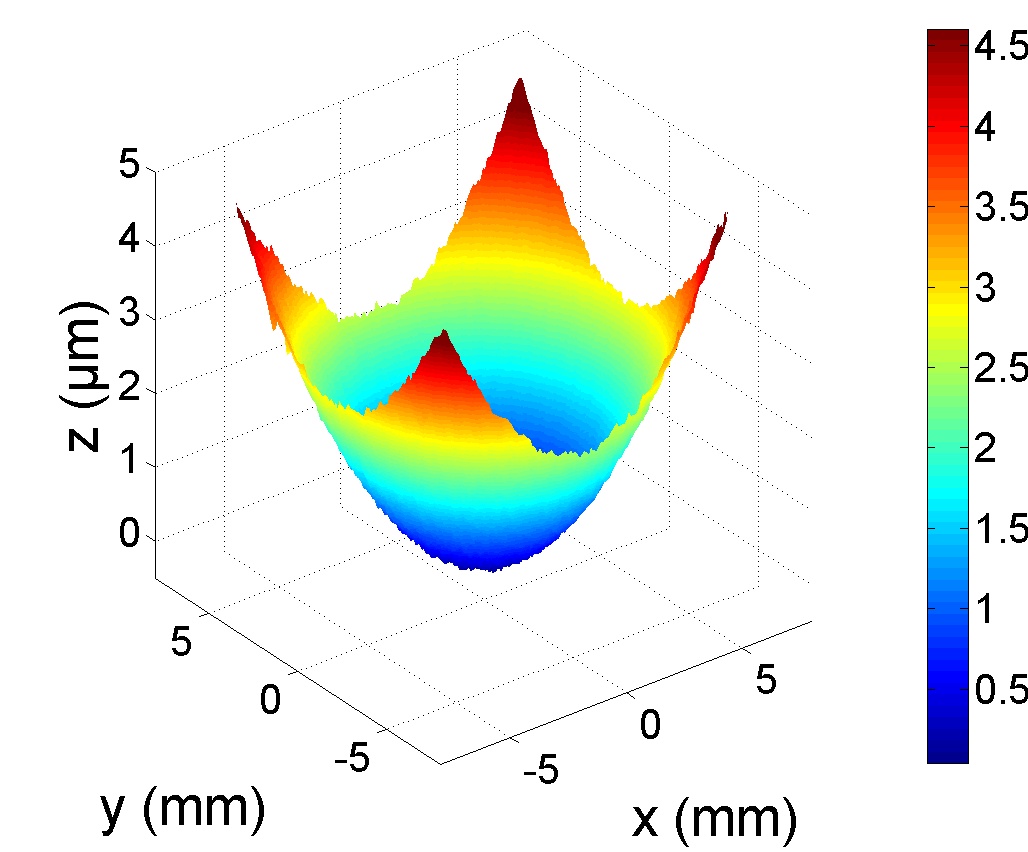}\hspace{0.001\textwidth}
\includegraphics[width=0.24\textwidth]{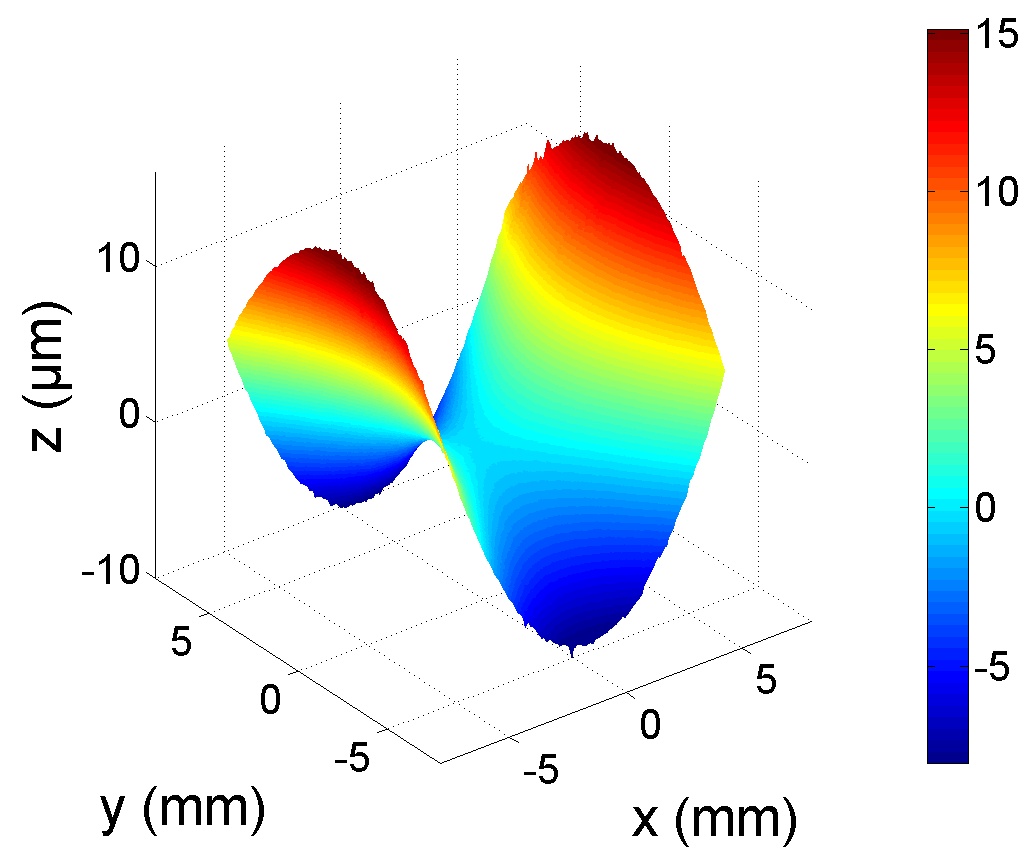}}\end{center}
\caption{Displacements at $625 \ \mathrm{V/mm}$ with single-sided electrodes at $320\, \upmu \mathrm{m}$ electrode spacing of (left to right): DIDE actuators with $s_{xx}\simeq 0$, $s_{yy}\simeq 0$, $s_{xx}\simeq s_{yy}$ and conventional in-plane polarized bending actuator (IDE only).}\label{3dplots}
\end{figure}
To test the predictions for the shape of the deformation and the amplitude shown in fig. \ref{lengthplot}, we produced single-sided configurations with $d=320 \, \upmu \mathrm{m}$ for $s_{yy} \simeq 0$ ($l= 443 \, \upmu \mathrm{m}$), $s_{xx} \simeq s_{yy}$ ($l= 620 \, \upmu \mathrm{m}$) and $s_{xx} \simeq 0$ ($l= 834 \, \upmu \mathrm{m}$), as well as a conventional IDE layout as the limiting case $l=0$. 
The displacements of these samples are shown in fig. \ref{3dplots}, where we clearly see the unidirectional, spherical and saddle-shaped bending profiles. In the unidirectional profiles, we also see the level of deviation from the design-profiles.

In order to also test different electrode spacings, we further produced configurations with $s_{xx}\simeq s_{yy}$ for single-sided electrodes at $d=720 \, \upmu \mathrm{m}$ ($l= 1183 \, \upmu \mathrm{m}$) and to test a different (half as thick) effective piezo thickness, we also produced double-sided electrodes at $d = 320 \, \upmu \mathrm{m}$ and $d=720 \, \upmu \mathrm{m}$ ($l= 530 \, \upmu \mathrm{m}$ and $l= 1078 \, \upmu \mathrm{m}$). As a reference, we compare our actuators also with a conventional transversely polarized (``$d_{31}$'') actuator prepared with the same materials and the same dimensions. The only differences compared to the in-plane polarized piezos are that the electrodes are not structured and that the initial transverse polarization was kept, so the only processing was to cut and glue the piezo.

\begin{figure}
\begin{center}{\includegraphics[width=0.49\textwidth]{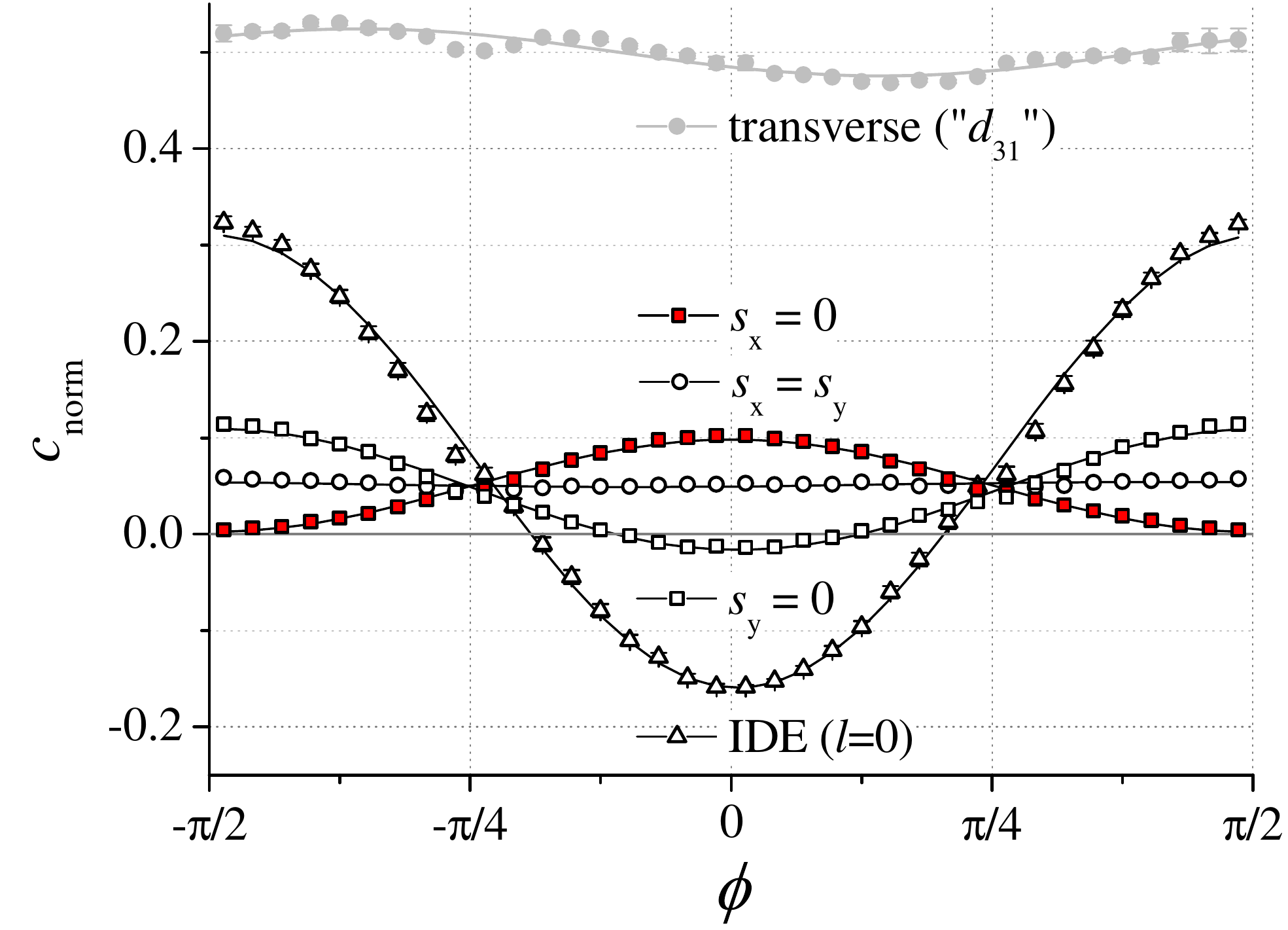}}\end{center}
\caption{The curvature $c(\phi)$ for the displacement profiles shown in fig. \ref{3dplots}, compared to the fit. The curves are obtained at the same field strength and $c_{norm}$ is normalized such that the average of the curve of the $d_{31}$ actuator would be $-\frac{1}{2}$, but for $d_{31}$, we changed the sign for the ease of plotting. For clarity, the error bars are not shown for the sphere and the unidirectional actuators, where they are smaller than or of the order of the symbol size.}\label{ccurve}
\end{figure}
To analyze the results, we described the actuator surface in polar coordinates, $z(r,\phi)$, which we let run, however, over $r\in [-|r_{min}|,r_{max}]$ and $\phi\in [-\pi/2,\pi/2]$. We performed quadratic fits in the radial direction for different angles around the center of the actuators, 
\begin{equation}\label{squarefit}
z(r,\phi)\  =\  a(\phi) + b(\phi) r + c(\phi) r^2
\end{equation} where the offset $a$ and tilt $b$ are not of interest.
Then, we fit a harmonic angular dependence with an alignment offset  $\delta$,
\begin{equation}\label{sinangle}
c(\phi) = \alpha + \beta \cos (2 \phi + \delta) \ 
\end{equation}
as shown in fig. \ref{ccurve}. There, we see that the fit parametrizes the displacement very well.
$\alpha$ and $\beta$ then give us a measure for the displacements $s_{xx}$ and $s_{yy}$ which is independent of any misalignments, as by comparision of equations (\ref{squarefit}) and (\ref{sinangle}) with equation (\ref{bendeq}), we get
\begin{equation}\label{alphas}
\alpha \, = \, - \frac{s_{xx}+s_{yy}}{4n} \ \ \mathrm{and} \ \ \beta \, = \, \frac{s_{yy}-s_{xx}}{4n} \ .
\end{equation}

To compare the measurements to the predictions of section \ref{simsec}, we can first of all identify from eq. (\ref{alphas}) $\gamma := \frac{\alpha \mp \beta }{\alpha \pm \beta}$ with $\frac{s_{xx,yy}}{s_{yy,xx}}$. As in the real, non-linear system, this ratio may still depend on the voltage, we compare the error-weighted mean over the voltage, $\left< \gamma \right>_U$ to the strain ratio. The RMS mean deviation of this distribution, $\sigma_\gamma$ may give an indication of the effects due to non-linearity and hysteresis of the piezo material.

\begin{figure}
\begin{center}{{\includegraphics[width=0.48\textwidth]{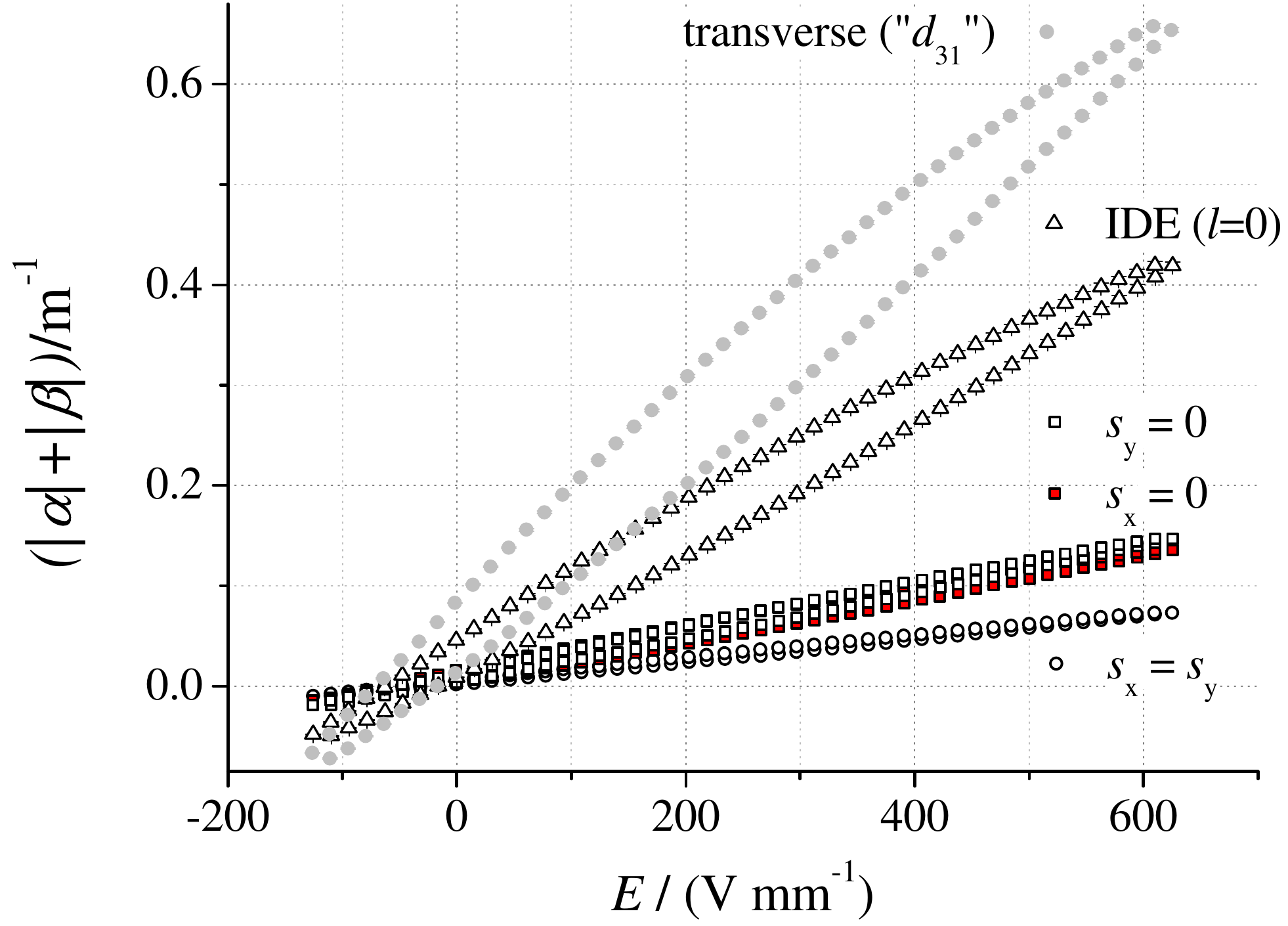}\hspace{0.02\textwidth}\includegraphics[width=0.48\textwidth]{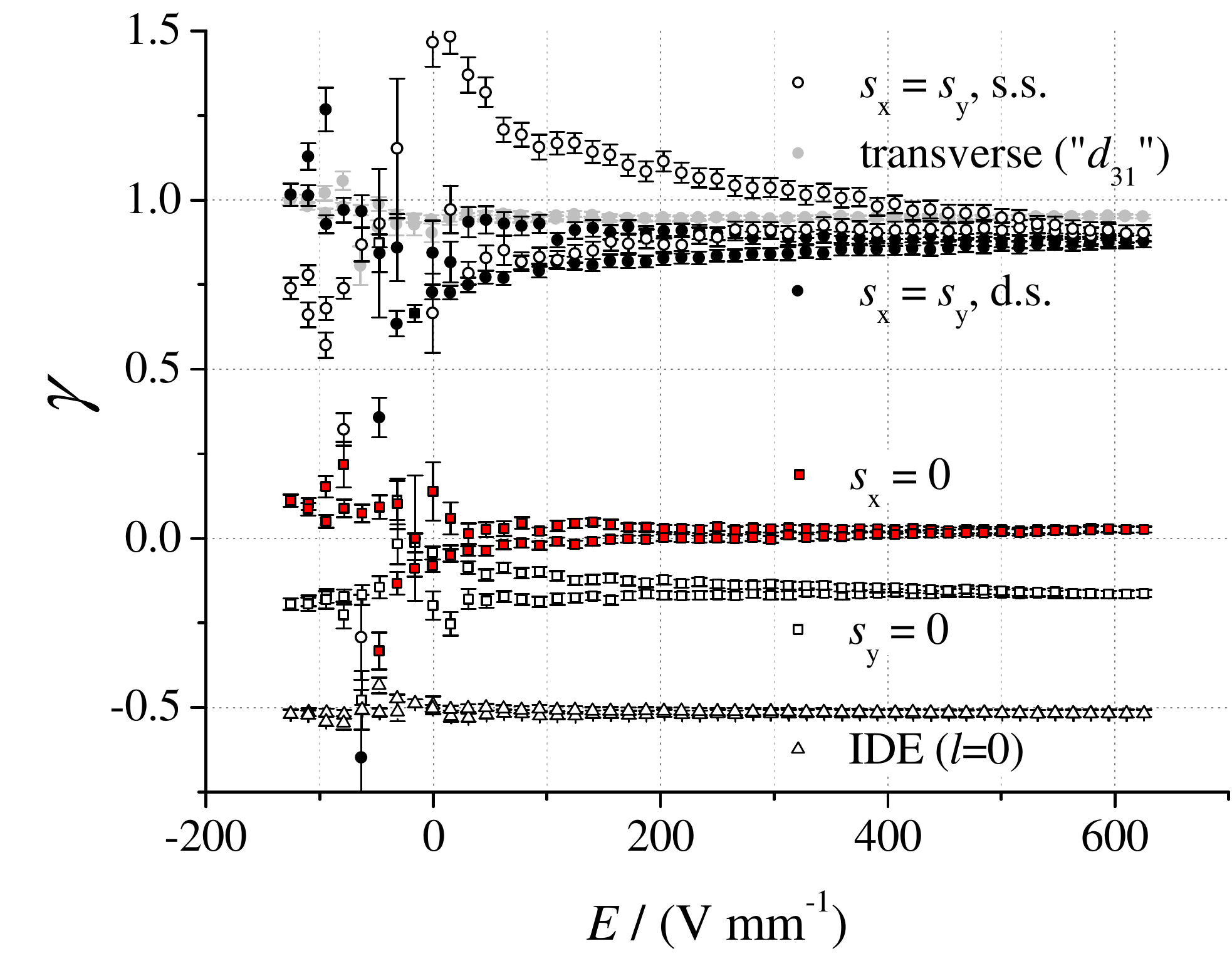}}}\end{center}
\caption{Left: Total displacement of the bending profiles shown in fig. \ref{3dplots}, as a function of the applied electric field. Strictly speaking, we do not show $|\alpha|+|\beta|$ but $\pm \alpha \pm \beta$ such that it matches $|\alpha|+|\beta|$ at the maximum field strength. Right: The ratio of curvatures $\gamma$. Here, we show in addition the curve for the $720 \, \upmu \mathrm{m}$ spaced, double-sided electrodes. The error bars are not shown for large field strengths, where they become smaller than the symbol sizes.}\label{parplots}
\end{figure}

In section \ref{simsec}, we used the larger of $|s_{xx}|$ and $|s_{yy}|$ as a measure for the displacement and took it relative to $s_{max} = d_{33}|E|$. From eq. (\ref{sinangle}), we see that the largest curvature occurs either when $2 \phi + \delta = 0 $ or $2 \phi + \delta = \pi $ and is given in magnitude by $|\alpha| + |\beta|$. In contrast to the ideal calculation in section \ref{theory}, the piezo material has a non-linear behavior with hysteresis. To consistently extract information about the displacement, we then take the mean of both branches of the Rayleigh loop at a given nominal electric field strength, and then fit a third oder polynomial
\begin{equation}
|\alpha| + |\beta| = \eta_0 + \eta_1 E + \eta_2 E^2 + \eta_3 E^3 \ .
\end{equation}
We then use $\eta_1$ as a measure for the displacement. Using again $d_{31}\simeq - \frac{1}{2} d_{33}$, we can use the curvature of the $d_{31}$ actuator as a normalization, like we used $s_{max}$ in section \ref{simsec}, so we define 
\begin{equation}
\tilde \eta = \eta_1/(2 \eta_1^{(d31)})
\end{equation}
as the parameter to compare with $\frac{s_{xx,yy}}{s_{max}}$ -- in the linear case and when $d_{31}= - \frac{1}{2} d_{33}$ they are equal by eq. (\ref{alphas}). In fig. \ref{parplots}, we show $|\alpha| + |\beta|$  and $\gamma$.
%
The noise in $\gamma$ is significant at small displacements, where the denominator of the fraction becomes small. There, we also observe a significant non-linear behavior, most prominently for the spherical displacements, $s_{xx} = s_{yy}$. This effect is, however, much smaller in the wider electrode spacings and single-sided configurations as seen in table \ref{restab}. We suppose that this is due to the large field near the electrodes as we discussed in section \ref{theory} and this region contributes more in the narrowly-spaced, single-sided setups than the widely-spaced double-sided setups. As seen in fig. \ref{lengthplot}, $s_{xx} = s_{yy}$ is most sensitive to variations of the geometry and thus probably also to nonlinearities.

\begin{figure}
\begin{center}{\footnotesize
\begin{tabular}{|r|l|c|c|c|c|@{\hspace{2pt}}c@{\hspace{2pt}}|@{\hspace{2pt}}c@{\hspace{2pt}}|c|c|c|}
\hline
 & $\frac{l}{d+t}$ & $h$ &  $\frac{d}{h}$ & $\frac{d}{t}$ & $\frac{t}{h}$ & $\left< \gamma \right>_{theory}$ & $\left<\gamma\right>$ & $\sigma_\gamma$ & $\tilde \eta_{theory} $ & $ \tilde \eta$ \\
\hline
\hline
conventional $d_{31}$ & - & $120\, \upmu \mathrm{m}$  & - & - & - &  1 & $ 0.950(1)$ & $0.0022(6)$ & 1/2 & 1/2 \\
\hline
IDE & 0 & $120 \upmu \mathrm{m}$ & 8/3 & 4 & 2/3 & -0.72(1) &  $-0.514(2)$ & $0.0000(2)$ & 0.466(7) & $ 0.3198(3)$ \\
\hline
DIDE $s_{yy}\simeq 0$ & 1.11 & $120\, \upmu \mathrm{m}$ & 8/3 & 4 & 2/3 & -0.18(7) & $-0.154(2)$ & $0.0016(8)$ & 0.155(13) & $ 0.1115(3)$ \\
\hline
$s_{xx}\simeq 0$ & 2.09 & $120\, \upmu \mathrm{m}$ & 8/3 & 4 & 2/3 & -0.17(4)  & $0.017(2) $ & $0.008(3) $ & 0.142(5) & $ 0.1036(2)$ \\
\hline
$s_{yy}\simeq s_{xx}$: s.s. & 1.55 & $120\, \upmu \mathrm{m}$ & 8/3 & 4 & 2/3 & 0.85(23)  & $0.936(5) $ & $ 0.039(7) $ & 0.064(9) & $0.0556(2) $ \\
\hline
double-sided & 1.33 & $60\, \upmu \mathrm{m}$ & 16/3 & 4 & 4/3 & 0.84(19)  & $0.860(4) $ & $ 0.02(1)$ & 0.108(10) & $ 0.0806(2) $ \\
\hline
 $720\, \upmu \mathrm{m}$ s.s. & 1.48 & $120\, \upmu \mathrm{m}$ & 6 & 9 & 2/3 & 0.87(10)  & $0.587(3) $ & $ 0.003(2) $ & 0.112(6) & $ 0.1166(2)$ \\
\hline
$720\, \upmu \mathrm{m}$ d.s. & 1.35 & $60\, \upmu \mathrm{m}$ & 12 & 9 & 4/3 & 0.88(10)  & $ 0.720(2) $ & $0.0005(5)$ & 0.140(8) & $ 0.1259(2) $ \\
\hline
\end{tabular}}\end{center}
\caption{Values for the displacements of the various demonstrators, errors in the last (or last two) digit(s) are written in parenthesis. The expected values of $\frac{s_{xx,yy}}{s_{yy,xx}}$ do not exactly match the labelling as the prototypes were produced according to an earlier, slightly different, model. The uncertainty value for the theory prediction corresponds to a $\pm 20 \upmu \mathrm{m}$ uncertainty in the finger length $l$.}\label{restab}
\end{figure}

Table \ref{restab} shows the experimental results, and they are furthermore indicated in fig. \ref{lengthplot}.  The  value $\gamma = 0.950\pm 0.001$ of the transversely polarized actuator serves as an indicator for the anisotropy in the material and the variations in the processing and assembly, as in theory the $d_{31}$ should give $\gamma = \frac{s_{xx,yy}}{s_{yy,xx}} = 1$. While the deviations from the predictions are significant compared to the measurement uncertainties, they are still within the range that is expected from
the uncertainties in the processes. One uncertainty contribution that we have included in the theory expectation values as an example is the structuring accuracy of the electrodes around $10 \ldots 20 \, \upmu \mathrm{m}$ due to the laser spot size, scanner linearity and etching in the $\mathrm{HNO}_3$ cleaning. Geometrically, there is also the thickness variation of the piezo layer of $10 \, \upmu \mathrm{m}$ with roughly another $10 \, \upmu \mathrm{m}$ from the laser-ablation of the electrodes, which may also damage the piezo material below the surface due to heat or temperature gradients. During the cleaning process, partial detachment of the electrodes from the ceramic may occur or the porous ceramic may be affected by the $\mathrm{HNO}_3$, which we observed to attack the PZT when etched at larger temperatures around $50^\circ \mathrm{C}$ for several minutes. 
An other leading error source may be the edge effects, both mechanically due to a non-perfect alignment of the glass layer and a not perfectly square shape and electrically due to the narrow rim of passive piezo material below and outside the outer electrode and due to the outer electrode itself. This may have a particularly strong effect on the IDE actuator ($l=0$), which would otherwise see no expansion along the electrodes but now has around $10\%$ of its area polarized in this direction. Finally, there is also the assumption  $d_{33} = -2 d_{31}$ that typically deviates around $10\%$ from the actual ratio and may give another significant error source. What we have also completely ignored is the variation of the strain over the thickness of the piezo, i.e. over the $z$ direction. This will also contribute some bending effect.

A systematic effect that we see is the deviation of $\tilde{\eta}$, as the displacements of the DIDE actuators are slightly smaller than the predictions. One particular reason for this may be the fact that the DIDE actuators have been thermally de-polarized and then electrically re-polarized while the reference out-of-plane actuator kept its initial polarization from the manufacturing process. It is not unlikely that our process significantly reduces the piezo coefficients. To test this, we check the consistency by plotting $\tilde \eta$ in fig. \ref{lengthplot} and find that they are consistent within what we expect from the processing accuracies. The single sided layouts seem to be closer to the predictions than the double-sided layouts.
\section{Summary and Conclusions}\label{conclusions}
We have successfully demonstrated that the characteristic anisotropic property of piezoelectric materials can be used to create actuators with tunable anisotropy of the deformations. 

In section \ref{theory}, we found in a material- and geometry-independent parametrization that the adjustment of the bending ratio is bought at the expense of a reduced displacement and that ellipsoidal displacements will always be smaller than the displacement of an isotropic spherical transversely (out-of-plane) polarized actuator. We discussed the efficiency of various doubly interdigitated electrode layouts, found an interesting scaling property with the film thickness and showed that $80-90\%$ of the optimal DIDE design or $50-80\%$ of a hypothetical optimal electrode design can be realistically achieved.

To test our model, we used a straightforward rapid-prototyping process, described in sec. \ref{process} to produce demonstrators with various displacement ratios and different types of electrode layouts. The analysis of the results of the profilometry in sec. \ref{characterization} was performed insensitive to misalignments and local defects. Overall, the results demonstrated that the method was successful in producing the desired shapes of the displacements. While the magnitude of the displacements of the in-plane polarized films was overall smaller than predicted, they are close to the predictions when compared relative to each other. The overall deviation may be due to the electrode structuring, varying film thickness, cleaning process and repolarization process. 
The deviation of the ratio of strains from the exact predictions was smaller than the deviation of the overall displacement, but it was still significant.
This is in some parts due to the simplifications in the theoretical model and in other parts due to large uncertainties in the fabrication process. These deviations can be taken 
care of by a more detailed model and more accurate processing, up to some remaining effects due to nonlinearity and saturation.
%
%
%
%
%

The principle that we demonstrated here may be extended to dynamically adjustable bending ratios, and these actuators may be applied to various fields. We will demonstrate this, and an application to adaptive optics in an upcoming paper \cite{omemsmoritz}. It would be interesting to see how this method can be applied to the direct piezoelectric effect to produce sensors that are selectively sensitive to specific strains and stresses or to specific resonance modes; and whether it is useful in other fields, for example in electroactive polymers.

\section*{Acknowledgments}
The research of M.W. is financed by the Baden-W\"urttemberg Stiftung gGmbH under the project ``ADOPT-TOMO'',
the research of M.S. is supported by DFG grant WA 1657/1-2
and the research of J.B. by DFG grant WA 1657/3-1.
%
\appendix
\section{Derivation of the mean strain}\label{deriv}
In this appendix, we derive the mean strain of the piezo sheets with a DIDE structure according to the principle described in section \ref{basic}.
Let us split each unit cell of the DIDE structure into a large numer of smaller segments, denoted by subscript $n$ with area $a_n$, in which the electric field created by the electrodes is homogeneous, $\vec{E}_n$. In each such segment, we can define a coordinate system $\left(x^{(n)},y^{(n)},z^{(n)} \right)$ that is aligned with the electric field, for example such that $\vec{E}_n = E \hat{x}^{(n)}$. As the piezo material is polarized with the same electrodes that are used for the driving field, the polarization is then also in the $x^{(n)}$ direction, and the strain tensor in this segment in the rotated coordinate system is hence given in component form (denoted by underlining) by
\begin{equation}
\underline{\underline{s}}_n^{(n)} \, = \, \begin{pmatrix}d_{33} & 0 & 0 \\ 0 & d_{31} & 0 \\ 0 & 0 & d_{31} \end{pmatrix} \ . 
\end{equation}
Now, we rotate back this coordinate system into the original coordinate system $(x,y,z)$ of the piezo. This can be done with a suitable rotation matrix $\underline{\underline{M}}^{(n)}$ that rotates 
\begin{equation}
\underline{E}_n \, = \, \underline{\underline{M}}^{(n)} \cdot \underline{E}_n^{(n)} \ \ \mathrm{and} \ \ 
\underline{\underline{s}}_n  \, = \, \underline{\underline{M}}^{(n)} \cdot \underline{\underline{s}}_n^{(n)} \cdot \left({\underline{\underline{M}}^{(n)}}\right)^t \ ,
\end{equation}
where we kept in mind that rotation matrices are orthogonal, i.e. ${\underline{\underline{M}}}^{-1} \, = \, {\underline{\underline{M}}}^t$ .
Finally, one has to take the average over a unit cell (or if necessary over two neigbouring mirror symmetric elements that we falsely call a ``unit cell'' as only two of them together form a real unit cell), such that 
\begin{equation}
\underline{\underline{s}} \, = \, \frac{\sum_n a_n \underline{\underline{s}}_n}{\sum_n a_n } \, =: \,  \left<\underline{\underline{s}} \right> \ .
\end{equation}
Obviously, the averaging applies only to the in-plane components $x$ and $y$ that we are interested in.

For simplicity, we first look at the two-dimensional system in the limiting case of an infinitely thin piezo sheet. In this case, the rotation matrix may be written as 
\begin{equation}
\underline{\underline{M}}^{(n)} \, = \, \begin{pmatrix} \cos \varphi_n & -\sin \varphi \\ \sin \varphi_n & \cos \varphi \end{pmatrix} \ ,
\end{equation}
such that the electric field and strain tensor are given by
\begin{eqnarray}
\underline{E}_n & = & E \begin{pmatrix} \cos \varphi_n \\ \sin \varphi_n \end{pmatrix}  \ \ \mathrm{and} \\ \nonumber  \underline{\underline{s}}_n & = & E \begin{pmatrix} d_{33} \cos^2 \varphi_n + d_{31} \sin^2 \varphi_n & (d_{33} - d_{31})\sin\varphi_n \cos\varphi_n \\ (d_{33} - d_{31})\sin\varphi_n \cos\varphi_n & d_{33} \sin^2 \varphi_n + d_{31} \cos^2 \varphi_n \end{pmatrix} \ .
\end{eqnarray}
Looking at fig. \ref{layout}, we see that neigbouring cells are symmetric under $\varphi \rightarrow -\varphi$. Hence, an average over the off-diagonal components vanishes and we are left with a diagonal strain tensor. Noting also that $\sin\varphi \, = \, \frac{E_{n,y}}{E}$ and $\cos\varphi \, = \, \frac{E_{n,x}}{E}$, we obtain
\begin{equation}\hspace{-2.5cm}
s_{xx}  =  d_{33}\left<\frac{E_x^2}{|E|}\right>+d_{31}\left<\frac{E_y^2}{|E|}\right> \ , \ \ 
s_{yy}  =  d_{33}\left<\frac{E_y^2}{|E|}\right>+d_{31}\left<\frac{E_x^2}{|E|}\right>  \ \mathrm{and} \ \ 
s_{xy}  = s_{yx}  = 0
\end{equation}

Finally, let us look at the full three-dimensional case. The rotation group in three dimensions, $SO(3)$, can be represented in different ways and has three degrees of freedom. One can visualize these by first mapping one point on a unit sphere to another point (two d.o.f.) and then rotating the coordinate system around the corresponding unit vector.
In the three dimensional real representation, any rotation can be generated by subsequent rotations around the three directions,
\begin{eqnarray}\nonumber
\underline{\underline{M}}_x  =  \begin{pmatrix} 1 & 0 & 0 \\ 0 & \cos \varphi_x & -\sin \varphi_x \\ 0 & \sin \varphi_x & \cos \varphi_x \end{pmatrix} \ , \ \ 
\underline{\underline{M}}_y \, = \, \begin{pmatrix} \cos \varphi_y & 0 & \sin \varphi_y \\ 0 & 1 & 0 \\ -\sin \varphi_y & 0 & \cos \varphi_y \end{pmatrix} \ \mathrm{and} \\
\underline{\underline{M}}_z  =  \begin{pmatrix} \cos \varphi_z & -\sin \varphi_z & 0 \\ \sin \varphi_z & \cos \varphi_z &0 \\ 0 & 0 & 1 \end{pmatrix} \ .
\end{eqnarray}
First, we map the components of the electric field to the ``original'' coordinate system of the DIDE structure. As this is only the rotation of a vector, it has only two degrees of freedom and the third degree of freedom is a degeneracy, that can be visualized as a rotation around the axis of the field vector. The possible rotations include $\underline{\underline{M}}_x^{(n)} \cdot \underline{\underline{M}}_y^{(n)} \cdot \underline{E}_n^{(n)}$, $\underline{\underline{M}}_x^{(n)} \cdot \underline{\underline{M}}_z^{(n)} \cdot \underline{E}_n^{(n)}$,
$\underline{\underline{M}}_y^{(n)} \cdot \underline{\underline{M}}_z^{(n)} \cdot \underline{E}_n^{(n)}$ and 
$\underline{\underline{M}}_z^{(n)}\cdot  \underline{\underline{M}}_y^{(n)} \cdot \underline{E}_n^{(n)}$. Rotating first around $\hat{x}^{(n)}$, however, would be trivial. When we rotate a tensor, these rotations are in general not equivalent anymore but in our case the rotational symmetry in the $\left(y^{(n)},z^{(n)} \right)$ plane ensures that they are also equivalent for tensors. Taking the latter rotation, we obtain the rotation matrix
\begin{equation}
\hspace{-1cm}\underline{\underline{M}}^{(n)} \, = \, \underline{\underline{M}}_z^{(n)}\cdot  \underline{\underline{M}}_y^{(n)} \, = \,
\begin{pmatrix}
 \cos\varphi_{y,n} \cos\varphi_{z,n} & -\sin\varphi_{z,n}  & \sin\varphi_{y,n} \cos\varphi_{z,n} \\
 \cos\varphi_{y,n} \sin\varphi_{z,n} & \cos\varphi_{z,n} & \sin\varphi_{y,n} \sin\varphi_{z,n} \\
 -\sin\varphi_{y,n} & 0 & \cos\varphi_{y,n}
\end{pmatrix}\ .
\end{equation}
Now, the electric field and strain are given by
\begin{eqnarray}
\underline{E}_n  = \, E \begin{pmatrix} \cos\varphi_{y,n} \cos\varphi_{z,n} \\ \cos\varphi_{y,n} \sin\varphi_{z,n} \\ -\sin\varphi_{y,n} \end{pmatrix}  \ \ \mathrm{and} \\   \nonumber
\underline{\underline{s}}_n  = \,   \\ \nonumber  \hspace{-3.5cm} { \scriptstyle E { \begin{pmatrix} \scriptstyle
 d_{33}\mcos^2\varphi_{y,n} \mcos^2\varphi_{z,n} + d_{31} \msin^2\varphi_{y,n} \mcos^2\varphi_{z,n} + d_{31} \msin^2\varphi_{z,n} & \scriptstyle
\msin\varphi_{z,n}\mcos\varphi_{z,n}\left(d_{33} \mcos^2\varphi_{y,n} - d_{31} \mcos^2\varphi_{y,n}  \right) & \scriptstyle
-\msin\varphi_{y,n}\mcos\varphi_{y,n}\mcos\varphi_{z,n}\left(d_{33} - d_{31} \right) \\\scriptstyle
\msin\varphi_{z,n}\mcos\varphi_{z,n}\left(d_{33} \mcos^2\varphi_{y,n} - d_{31} \mcos^2\varphi_{y,n}  \right) &\scriptstyle
 d_{33}\mcos^2\varphi_{y,n} \msin^2\varphi_{z,n} + d_{31} \msin^2\varphi_{y,n} \msin^2\varphi_{z,n} + d_{31} \mcos^2\varphi_{z,n} & \scriptstyle
-\msin\varphi_{y,n}\mcos\varphi_{y,n}\msin\varphi_{z,n}\left(d_{33} - d_{31} \right) \\ \scriptstyle
-\msin\varphi_{y,n}\mcos\varphi_{y,n}\mcos\varphi_{z,n}\left(d_{33} - d_{31} \right) & \scriptstyle
-\msin\varphi_{y,n}\mcos\varphi_{y,n}\msin\varphi_{z,n}\left(d_{33} - d_{31} \right) &\scriptstyle
d_{33} \msin^2  \varphi_{y,n} + d_{31} \mcos^2  \varphi_{y,n}
\end{pmatrix}}} \ ,
\end{eqnarray}
where $\msin$ stands for $\sin$ and $\cos$ has been abbreviated by $\mcos$.
As the rotation with $\underline{\underline{M}}_z^{(n)}$ axis is done last, it takes already place in the plane of the piezo sheet. Hence, the symmetry of the DIDE pattern under $\varphi_z \rightarrow -\varphi_z$ will again make the $x-y$ components vanish. After some straightforward trigonometry and replacing again the electric field components for the angles, we arrive with
\begin{eqnarray}
s_{xx}  =  d_{33}\left<\frac{E_x^2}{|E|}\right>+d_{31}\left<\frac{E_y^2}{|E|}\right>+d_{31}\left<\frac{E_z^2}{|E|}\right> \ , \\ \nonumber 
s_{yy}  = \, d_{33}\left<\frac{E_y^2}{|E|}\right>+d_{31}\left<\frac{E_x^2}{|E|}\right>+d_{31}\left<\frac{E_z^2}{|E|}\right>  \ \mathrm{and} \ \ \ 
s_{xy}  =  s_{yx} \, = \,0 \ .
\end{eqnarray}

\bibliography{piezobib}

\begin{thebibliography}{10}

\bibitem{astronomy}
Andreas Glindemann, Stefan Hippler, Thomas Berkefeld, and Wolfgang Hackenberg.
\newblock Adaptive optics on large telescopes.
\newblock {\em Experimental Astronomy}, 10(1):5--47, 2000.

\bibitem{piezoinject}
Alfred~LW Williams.
\newblock Piezoelectric fuel injector, July~13 1965.
\newblock US Patent 3,194,162.

\bibitem{thunderpat}
R.F. Hellbaum, R.G. Bryant, and R.L. Fox.
\newblock Thin layer composite unimorph ferroelectric driver and sensor, May~27
  1997.
\newblock US Patent 5,632,841.

\bibitem{thunderpaper}
K.M. Mossi and R.P. Bishop.
\newblock Characterization of different types of high performance thunder™
  actuators.
\newblock {\em Stainless Steel}, 302:0--1524, 1999.

\bibitem{lipca}
K.J. Yoon, S.~Shin, H.C. Park, and N.S. Goo.
\newblock Design and manufacture of a lightweight piezo-composite curved
  actuator.
\newblock {\em Smart Materials and Structures}, 11(1):163, 2002.

\bibitem{rainbow}
G.H. Haertling.
\newblock Rainbow ceramics- a new type of ultra-high-displacement actuator.
\newblock {\em American Ceramic Society Bulletin}, 73(1):93--96, 1994.

\bibitem{adlerpat}
R.~Adler.
\newblock Piezoelectric transducer and method for producing same, February~6
  1951.
\newblock US Patent 2,540,412.

\bibitem{SAW}
Herbert Matthews.
\newblock Surface wave filters: Design, construction, and use.
\newblock {\em New York, Wiley-Interscience, 1977. 521 p}, 1, 1977.

\bibitem{piezoefficiency}
W.~Beckert and W.S. Kreher.
\newblock Modelling piezoelectric modules with interdigitated electrode
  structures.
\newblock {\em Computational materials science}, 26:36--45, 2003.

\bibitem{piezofibers}
A.A. Bent, N.W. Hagood, and J.P. Rodgers.
\newblock Anisotropic actuation with piezoelectric fiber composites.
\newblock {\em Journal of Intelligent Material Systems and Structures},
  6(3):338--349, 1995.

\bibitem{flexbendmirr}
I.~Kanno, T.~Kunisawa, T.~Suzuki, and H.~Kotera.
\newblock Development of deformable mirror composed of piezoelectric thin films
  for adaptive optics.
\newblock {\em Selected Topics in Quantum Electronics, IEEE Journal of},
  13(2):155--161, 2007.

\bibitem{omemsmoritz}
Moritz St\"urmer, Matthias~C Wapler, Jens Brunne, and Ulrike Wallrabe.
\newblock Focusing mirror with tunable eccentricity.
\newblock In {\em Optical MEMS and Nanophotonics (OMN), 2013 International
  Conference on}. IEEE, 2013.

\end{thebibliography}
\bibliographystyle{unsrt}

\end{document}